\documentclass[reprint,preprintnumbers,nofootinbib,amsmath,amssymb,aps,prd,onecolumn]{revtex4-2}
\pdfoutput=1
\usepackage{amsmath}
\usepackage{amssymb}
\usepackage{booktabs}
\usepackage{hyperref}
\usepackage{graphicx}
\usepackage{multirow}
\usepackage{framed}
\usepackage{xcolor}

\hypersetup{
	colorlinks=true,
	linkcolor=blue,
	citecolor=magenta,
	filecolor=magenta,
	urlcolor=cyan
}

\renewcommand{\vec}[1]{\mathbf{#1}}
\newcommand{\unitvec}[1]{\hat{\mathbf{#1}}}

\begin{document}

\title{Identifying magnetic antiskyrmions while they form with convolutional neural networks}
\preprint{IPPP/22/18}

\author{Jack Y. Araz}
\email{jack.araz@durham.ac.uk}
\affiliation{Institute for Particle Physics Phenomenology, Durham University, South Road, Durham DH1 3LE, United Kingdom}

\author{Juan Carlos Criado}
\email{juan.c.criado@durham.ac.uk}
\affiliation{Institute for Particle Physics Phenomenology, Durham University, South Road, Durham DH1 3LE, United Kingdom}

\author{Michael Spannowsky}
\email{michael.spannowsky@durham.ac.uk}
\affiliation{Institute for Particle Physics Phenomenology, Durham University, South Road, Durham DH1 3LE, United Kingdom}

\begin{abstract}
  Chiral magnets have attracted a large amount of research interest in recent years because they support a variety of topological defects, such as skyrmions and bimerons, and allow for their observation and manipulation through several techniques. They also have a wide range of applications in the field of spintronics, particularly in developing new technologies for memory storage devices. However, the vast amount of data generated in these experimental and theoretical studies requires adequate tools, among which machine learning is crucial. We use a Convolutional Neural Network (CNN) to identify the relevant features in the thermodynamical phases of chiral magnets, including (anti-)skyrmions, bimerons, and helical and ferromagnetic states. We use a flexible multi-label classification framework that can correctly classify states in which different features and phases are mixed. We then train the CNN to predict the features of the final state from snapshots of intermediate states of a lattice Monte Carlo simulation. The trained model allows identifying the different phases reliably and early in the formation process. Thus, the CNN can significantly speed up the large-scale simulations for 3D materials that have been the bottleneck for quantitative studies so far. Moreover, this approach can be applied to the identification of mixed states and emerging features in real-world images of chiral magnets.
\end{abstract}

\maketitle

\section{Introduction}
\label{sec:intro}

Chiral magnets with Dzyaloshinskii-Moriya (DM) interactions~\cite{Dzyaloshinskii:1958,Moriya:1960zz} present a rich set of phases in which topologically non-trivial structures arise. Among them, one finds skyrmions~\cite{Skyrme:1962vh, Bogdanov:1989}, with unit topological charge; antiskyrmions, which are counterparts with opposite charge; and other objects with fractional charge, such as bimerons. These structures have been observed experimentally in a variety of materials~\cite{Muhlbauer:2009, Munzer:2009var, yu2010real, yu2011near, tokunaga2015new, woo2016observation, fujima2017thermodynamically, koshibae2016theory, hoffmann2017antiskyrmions, huang2017stabilization, camosi2018micromagnetics, kovalev2018skyrmions, bottcher2018b, jena2020elliptical, yu2021magnetic}. The interest in these objects goes beyond the determination of their fundamental properties, as they have applications in the field of spintronics~\cite{Fert:2013, tomasello2014strategy, song2020skyrmion, pinna2020reservoir, zazvorka2019thermal, huang2017stabilization}.

One of the critical theoretical tools for studying chiral magnets is the implementation of Monte Carlo simulations for a discretized version of them. It has been shown in Ref.~\cite{buhrandt2013skyrmion} that a 3D cubic spin lattice model with ferromagnetic and DM interactions can correctly reproduce the experimentally-determined phase diagram for materials that support Bloch skyrmions. An exploration of the consequences of varying both the strength and the internal structure of the DM interaction was performed in Ref.~\cite{Criado:2021gzp}. Changing their structure in this system generates Neel skyrmions and antiskyrmions, while the strength controls the size of the corresponding objects.

As these simulations produce more varied and complex states and are used to map larger parameter spaces, it becomes crucial to develop automatic methods for dealing with the data they generate. The supervised machine learning framework provides an excellent toolbox to do this. As long as they are adequately selected, machine learning models can be trained with a relatively small set of samples and then used to analyze any new data generated through simulations automatically. In particular, Convolutional Neural Networks (CNNs) have been successfully applied to this type of task: to identify the phase~\cite{PhysRevB.98.174411, salcedo2020deep, singh2019application, PhysRevB.105.214423, perzhu2020computer, kapitan2021numerical}, the topological charge~\cite{matthies2022topological} or the DM interaction~\cite{kawaguchi2021determination} from 2D images of a spin-lattice; and to find the phase from videos of the lattice~\cite{wang2021learning}.

In this work, we use a CNN in a multi-label approach to identify features such as skyrmions, bimerons, helical and ferromagnetic states, or hexagonal arrangements of the skyrmions. Several of these features can coexist in the same sample. The corresponding phase can then be inferred from the features. Furthermore, this approach accounts for states that mix two or more phases and allows for identifying features that may only appear in a small region of the image. We also apply this model to predict the features of the final state of a simulation from images of intermediate states. We then use it to construct a phase diagram by running the simulations for very short times and using the CNN to predict the final phase.

The rest of this paper is organized as follows. In Section~\ref{sec:theory} we review the Monte Carlo simulations that we perform for the generation of the training and test data. The data set itself, and the particular CNN that we use for it are described in Section~\ref{sec:method}. We show our results in Section~\ref{sec:results}, and give our conclusions in Section~\ref{sec:conclusions}.

\section{Simulations}
\label{sec:theory}

\begin{table}
  \centering
  \begin{tabular}{ccc}
    \toprule
    Label
    & Description
    & Examples \\
    \midrule
    \texttt{antiskyrmion}
    & \begin{minipage}{5cm} A circular region of spins anti-aligned with the extenal magnetic field. \end{minipage}
    & \raisebox{-0.5\totalheight}{
      \includegraphics[width=0.2\textwidth]{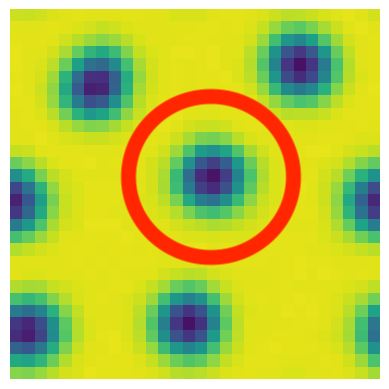}
      \includegraphics[width=0.2\textwidth]{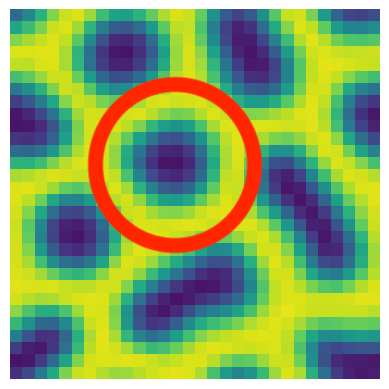}
      } \\
    \texttt{bimeron}
    & \begin{minipage}{5cm} A wall of spins (anti-aligned with the external magnetic field) that ends. \end{minipage}
    & \raisebox{-0.5\totalheight}{
      \includegraphics[width=0.2\textwidth]{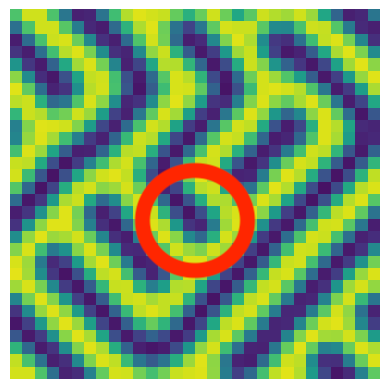}
      \includegraphics[width=0.2\textwidth]{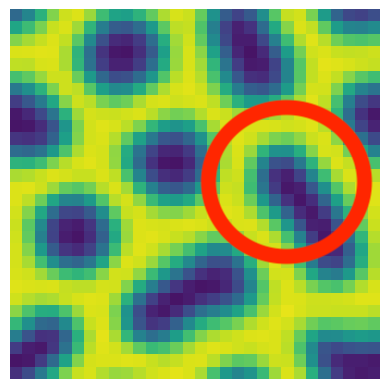}
      } \\
    \texttt{helix}
    & \begin{minipage}{5cm} At least two contiguous walls of spins anti-aligned with the external magnetic field. \end{minipage}
    & \raisebox{-0.5\totalheight}{
      \includegraphics[width=0.2\textwidth]{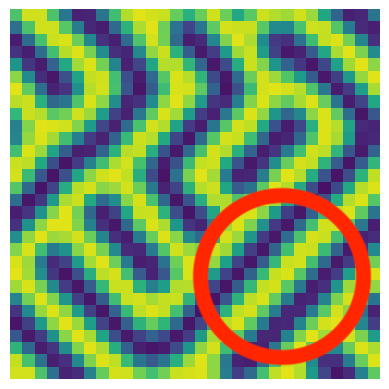}
      \includegraphics[width=0.2\textwidth]{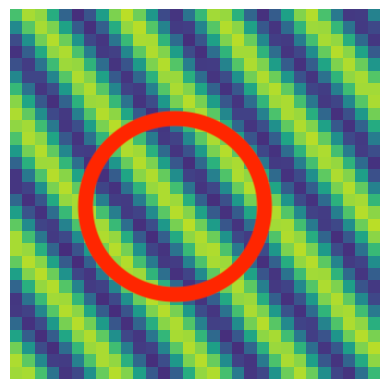}
      } \\
    \texttt{ferromagnetic}
    & \begin{minipage}{5cm}
      A region of spins aligned with the extenal magnetic field,
      either filling the full snapshot, or at least having a size
      larger than the typical distance between antiskyrmions.
    \end{minipage}
    & \raisebox{-0.5\totalheight}{
      \includegraphics[width=0.2\textwidth]{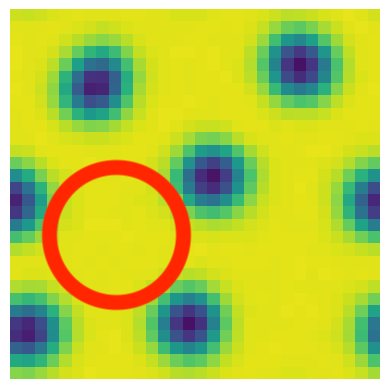}
      \includegraphics[width=0.2\textwidth]{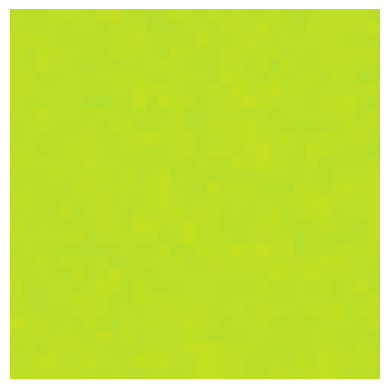}
      } \\
    \texttt{hexagonal}
    & \begin{minipage}{5cm} An arrangement of antiskyrmions forming a hexagonal lattice filling the entire snapshot. \end{minipage}
    & \raisebox{-0.5\totalheight}{
      \includegraphics[width=0.2\textwidth]{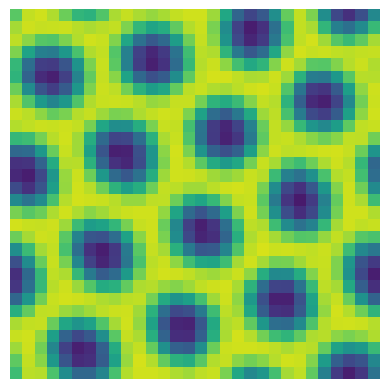}
      \includegraphics[width=0.2\textwidth]{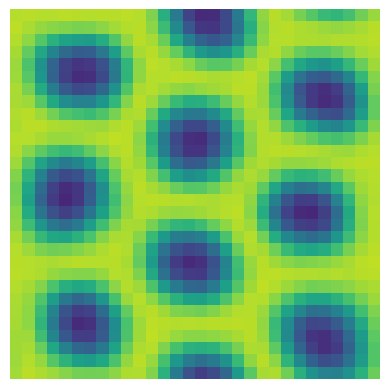}
      } \\
    \bottomrule
  \end{tabular}
  \caption{Labels used in the dataset. For each label (except \texttt{hexagonal}), if an object matching the description is observed, the corresponding element of the label vector is set to 1, otherwise it is set to 0.}
  \label{tab:labels}
\end{table}

The system we simulate is a chiral magnet with ferromagnetic and DM interactions. From a coarse-grained perspective, it can be described by a local continuum Hamiltonian for the magnetisation field. We discretise it in a 3D cubic spin-lattice of size $30 \times 30 \times 30$, with a unit vector $\vec{S}_{\vec{r}}$ at each lattice position $\vec{r}$ and periodic boundary conditions. To leading order in the lattice spacing, the Hamiltonian of this discrete system consists of the following nearest neighbours interactions:
\begin{align}
  H
 =
 - \sum_{\vec{r}} \Big\{
  J \; \mathbf{S}_{\vec{r}} \cdot \left(
   \mathbf{S}_{\vec{r} + \unitvec{x}}
   + \mathbf{S}_{\vec{r} + \unitvec{y}}
   + \vec{S}_{\vec{r} + \unitvec{z}}
   \right)
   + K \, \text{DM}_{\vec{r}}(\vec{S})
   + B \, \left(\vec{S}_{\vec{r}}\right)_z
   \Big\},
   \label{eq:hamiltonian}
\end{align}
where $\text{DM}_{\vec{r}}(\vec{S})$ is the DM interaction, $B$ is the magnitude of an external magnetic field applied along the $z$ direction, and $J$ and $K$ are parameters controlling the strength of the ferromagnetic and DM interactions, respectively. All these parameters of the lattice system are adimensional and proportional to the corresponding parameters of the physical system. Following Ref.~\cite{buhrandt2013skyrmion}, we correct the anisotropies generated by the finite lattice spacing by introducing next-to-nearest neighbours of the same form as the nearest-neighbours ones in Eq.~\eqref{eq:hamiltonian}, with adequate coefficients.

The DM interactions for different materials take different forms as functions of $\mathbf{S}$. Here, we select the structure that generates antiskyrmions, which is~\cite{Criado:2021gzp}:
\begin{equation}
  \text{DM}_{\vec{r}}(\vec{S})
  =
  (\vec{S}_{\vec{r}} \times \vec{S}_{\vec{r} + \unitvec{y}})_{x}
  - (\vec{S}_{\vec{r}} \times \vec{S}_{\vec{r} + \unitvec{x}})_{y}.
\end{equation}
When we extract data from the simulation to train the CNN, we will only use the $z$ component of $S$. Since Bloch/Neel skyrmions and antiskyrmions have similar distributions for the $z$ component, the choice we have made for the DM interaction does not lead to a loss of generality: the approach we use in Section~\ref{sec:method} would perform similarly for Bloch/Neel skyrmions. We check this numerically in Section~\ref{sec:results}.

At a finite temperature $T$, the probability of finding the system in a state with energy $E$ is proportional to $e^{-E/T}$. As for the other parameters, $T$ is here adimensional and proportionate to the physical temperature of the system. To reproduce this probability distribution in our simulation we use the Metropolis algorithm: the system is initialized in a random state and then updated iteratively by choosing a random spin and changing it to a new random direction with a probability
\begin{equation}
 p(\Delta E) = \min \left\{1, e^{-\Delta E / T}\right\},
\end{equation}
where $\Delta E$ is the energy difference between the system's energy after the potential change and the current one. To speed up the process, we divide the lattice into three non-interacting sublattices and update all spins in each sublattice in parallel, using a GPU, as described in Ref.~\cite{Criado:2021gzp}.

A rescaling of the parameters $J$, $K$, $B$ and $T$ by the same factor leaves the system invariant. We are thus free to fix the value of any of them without loss of generality. We pick $J = 1$. We set the remaining parameters $K$, $B$ and $T$ to constant values inside a given simulation but varying across different simulations.

We perform two types of procedures during a simulation, which we call \emph{themalization stages} and \emph{averaging stages}. A thermalization stage consists of 1000 lattice sweeps. A lattice sweep is an update of all of the spins in the lattice. An averaging step involves taking the average of 200 lattice configurations, with 50 sweeps performed between each consecutive pair. In a simulation, we perform thermalization and averaging stages alternatively, for a total of 20 each. The number of sweeps we take in both kinds of the stage is small compared to their typical size in other applications. As a result, the sequence of averaged configurations obtained from the averaging stages gives us a collection of \emph{snapshots} showing the dynamical evolution of the system as the final state is formed.

\section{Data preparation and training}
\label{sec:method}

\begin{figure}
  \centering
  \includegraphics[width=0.45\textwidth]{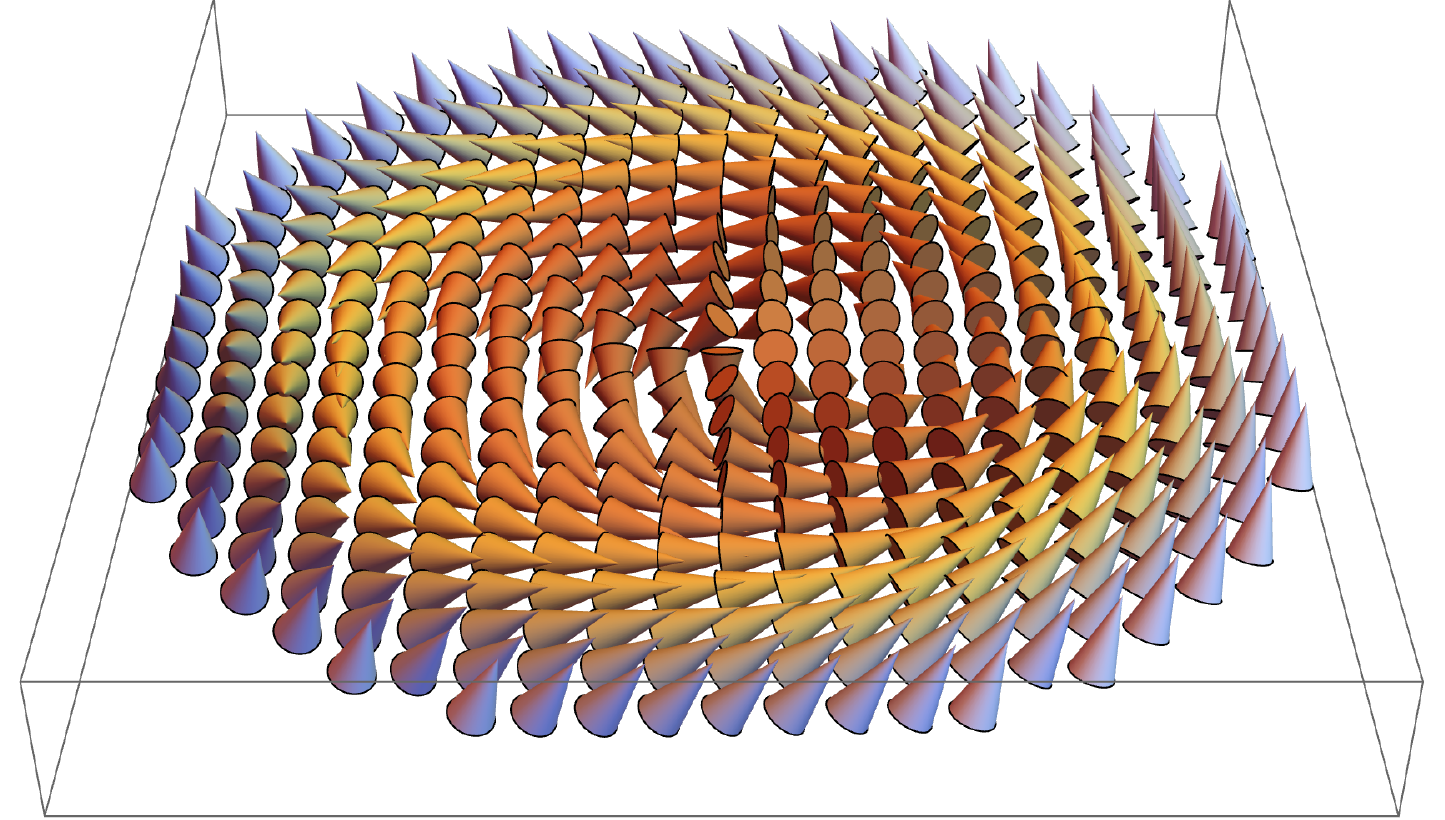} \qquad
  \includegraphics[width=0.45\textwidth]{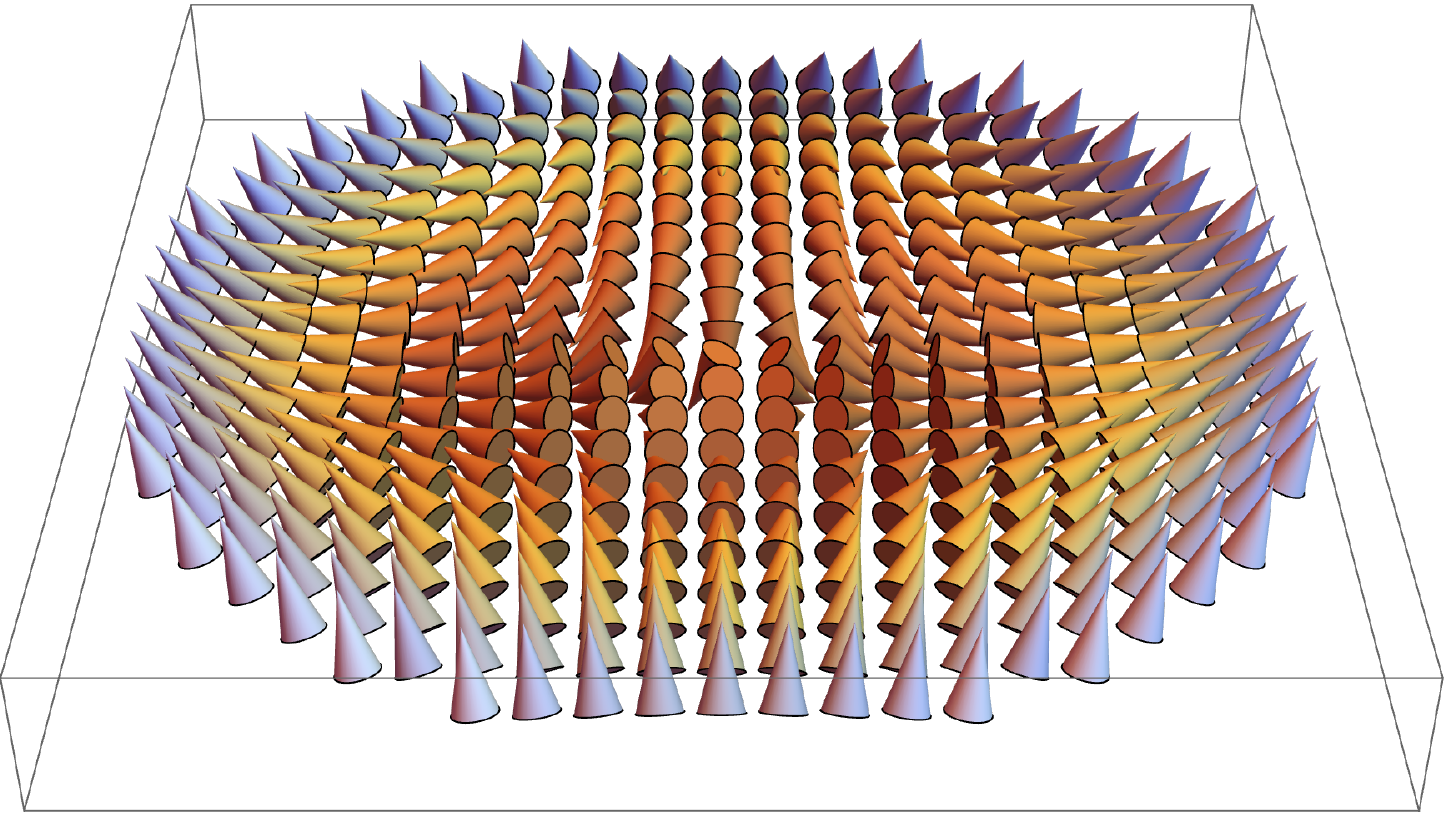}

  \vspace{20pt}

  \includegraphics[width=0.45\textwidth]{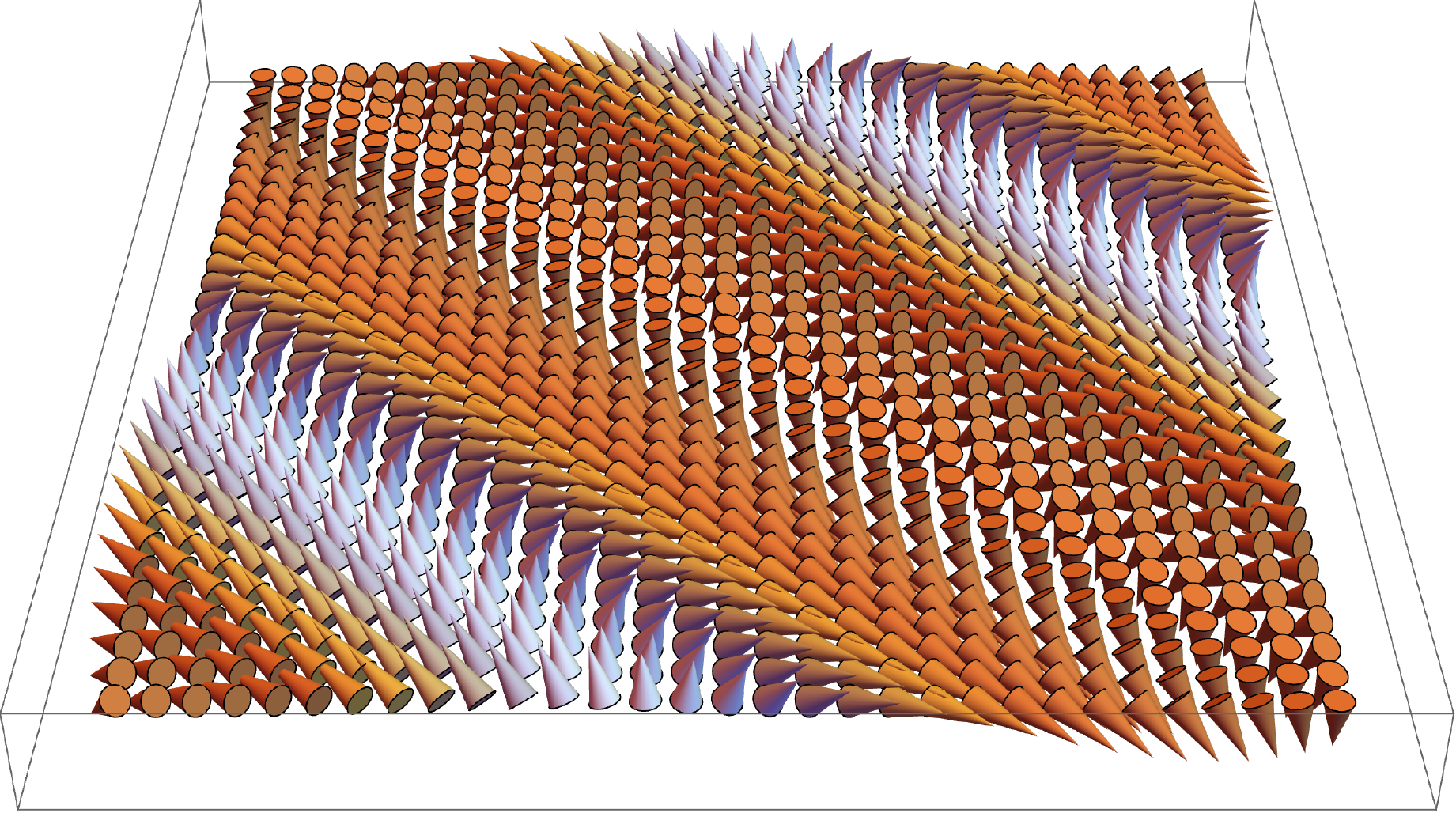} \qquad
  \includegraphics[width=0.45\textwidth]{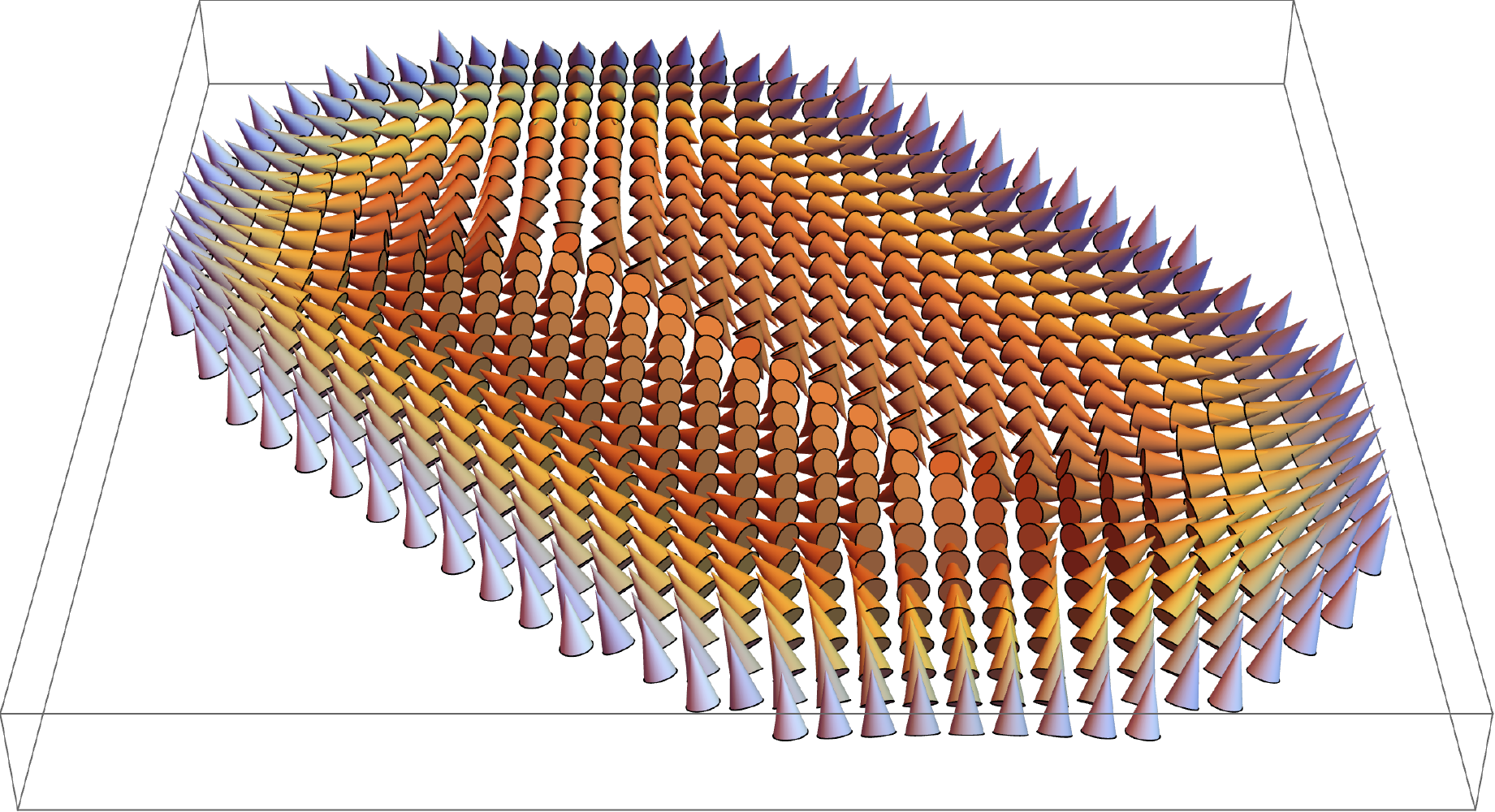}
  \caption{Spin arrangements for the different structures considered in this paper. Top left: Bloch skyrmion, with topological charge $Q=-1$ and helicity $\gamma = \pi/2$. Top right: antiskyrmion, with $Q = 1$ and $\gamma = 0$. Bottom left: helical state. Bottom right: bimeron, with the two end points having topological charge $Q = 1/2$.}
  \label{fig:illustration}
\end{figure}

To generate our data set, we run 1000 Monte Carlo simulations with values of $K$, $T$ and $B$ randomly chosen with a uniform distribution in the intervals $[0.7, 1.4]$, $[0.1, 1.3]$ and $[0.05, 0.35]$, respectively. As described in Section~\ref{sec:theory}, we obtain 20 equally-spaced 3D snapshots for each simulation. We perform an additional averaging stage over 2000 configurations for the final state. For each snapshot (and for the final state), we take a 2D slice of the lattice at constant $z = 0$ and keep only the $z$ component of the corresponding spins. This is enough information to identify the relevant objects in each case. We refer to the set of 20 2D snapshots plus the final 2D image (both of size $30 \times 30$) obtained in this way as a \emph{sample}.

To each sample, we attach a set of labels, chosen from the five ones presented in Table~\ref{tab:labels}, and representing the features observed in the final image. One or more of these labels can be assigned to the same sample. The spin arrangements corresponding to the \texttt{antiskyrmion}, \texttt{helix}, \texttt{bimeron} labels are illustrated in Fig.~\ref{fig:illustration}, together with the Bloch skyrmion configuration, which we consider briefly at the end of Section~\ref{sec:results}. Skyrmions and antiskyrmions are characterized by their topological charge~\cite{GOBEL20211}:
\begin{equation}
  Q = \frac{1}{4\pi} \sum_{\vec{r}} \vec{S}_{\vec{r}} \cdot (\vec{S}_{\vec{r} + \unitvec{x}}\times \vec{S}_{\vec{r} + \unitvec{y}}),
\end{equation}
which is $Q = -1$ for skyrmions and $Q = 1$ for antiskyrmions. As shown in Fig.~\ref{fig:illustration}, both share the same distribution for $\vec{S}_{\vec{z}}$. Different types of (anti-)skyrmions are differentiated by their vorticity $m$ and helicity $\gamma$. The are defined by the relation between two angles: the azimut $\Phi$ for any given spin $\vec{S}_{\vec{r}}$, and the polar angle $\phi$ for its position $\vec{r} - \vec{r}_0$ with respect to the center $\vec{r}_0$ of the structure. The defining relation is then:
  \begin{equation}
    \Phi = m \phi + \gamma.
  \end{equation}
  The vorticity $m$ is related in our case to the topological charge through $m = -Q$. The helicity might vary among configurations with the same topological charge. In particular, Bloch skyrmions have $\gamma = \pi/2$, while the antiskyrmions generated in our setup have $\gamma = 0$. The bimeron states are composed by two structures with $Q = 1/2$, with the arrangement of spins of one half of a $\gamma = 0$ antiskyrmion, connected by a wall-like domain as displayed in Fig.~\ref{fig:illustration} (see Refs.~\cite{PhysRevB.83.100408, rosales2022skyrmion}). The polarity, which characterizes the direction of the spins at the center of these configurations, is in all the cases we consider anti-aligned with the external magnetic field.

The set of labels from Table~\ref{tab:labels} is represented as a vector of 5 binary 0/1 components for each sample, which we call the \emph{label vector}. A component of the label vector being one means that the corresponding label is present, while a value of 0 indicates its absence. We remove 50 samples that do not present a clear structure with features matching any of the ones in Table~\ref{tab:labels}. We then standardise the entire data set via \texttt{RobustScaler} from the \texttt{scikit-learn} package.

We perform a separate training procedure for each snapshot index from 1 to 20. Among the 950 samples we have, we use 190 for testing purposes and 20\% of the remaining samples for validation. We augment the training data by a factor of 10 by applying a shift by a random amount in the horizontal and the vertical direction, wrapping around the edges. This has several related advantages. First, it increases the effective size of the data set by a factor of $30 \cdot 30$, the number of different shifts performed. Second, it enforces that the CNN learns the translational symmetry that the system possesses. Finally, it prevents over-fitting to particular characteristics localised at specific positions in the snapshots selected for training.

The network architecture has been formed via a simple convolution block followed by a fully connected layer. The convolution block has been constructed via a convolutional layer with 16 filters and four stride steps in each direction on the 2D image. The convolutional layer has been wrapped with a \texttt{ReLU} activation function, and its weights have been regularised via L2 with a weight of 0.01. A max-pooling layer has followed it, which takes the pixel with maximum impact within a square of $2\times 2$ pixels. The max-pooling layer has been sandwiched between two batch normalisation layers. The flattened output of the convolution block then has been fed into a fully connected layer with 16 nodes which, again, has been wrapped with \texttt{ReLU} activation function, and the weights have been regularised with the same L2 regulariser. Finally, the fully connected layer has been padded with dropout layers with 30\% probability. This network is designed to provide a five-dimensional output wrapped with a sigmoid activation function.

For training, we use the cross-entropy loss function, which we minimize using the \texttt{Adam} algorithm~\cite{Kingma2014AdamAM}, with a learning rate starting at $10^{-3}$ and decaying by a factor of 2 if the validation loss did not improve for 20 epochs. We perform a maximum of 200 training epochs for each snapshot, with the training being terminated if the validation loss function has not improved for 50 epochs. As mentioned above, to preserve the simplicity of the application, we used one architecture for all snapshots. However, since the last snapshots include more information than the initial snapshots, we observed overtraining within a short amount of epochs during the training of the earlier snapshots. Hence, we only used the network structure before it started to overtrain.

\section{Results}
\label{sec:results}

\begin{figure}
  \centering
  \includegraphics[scale=0.5]{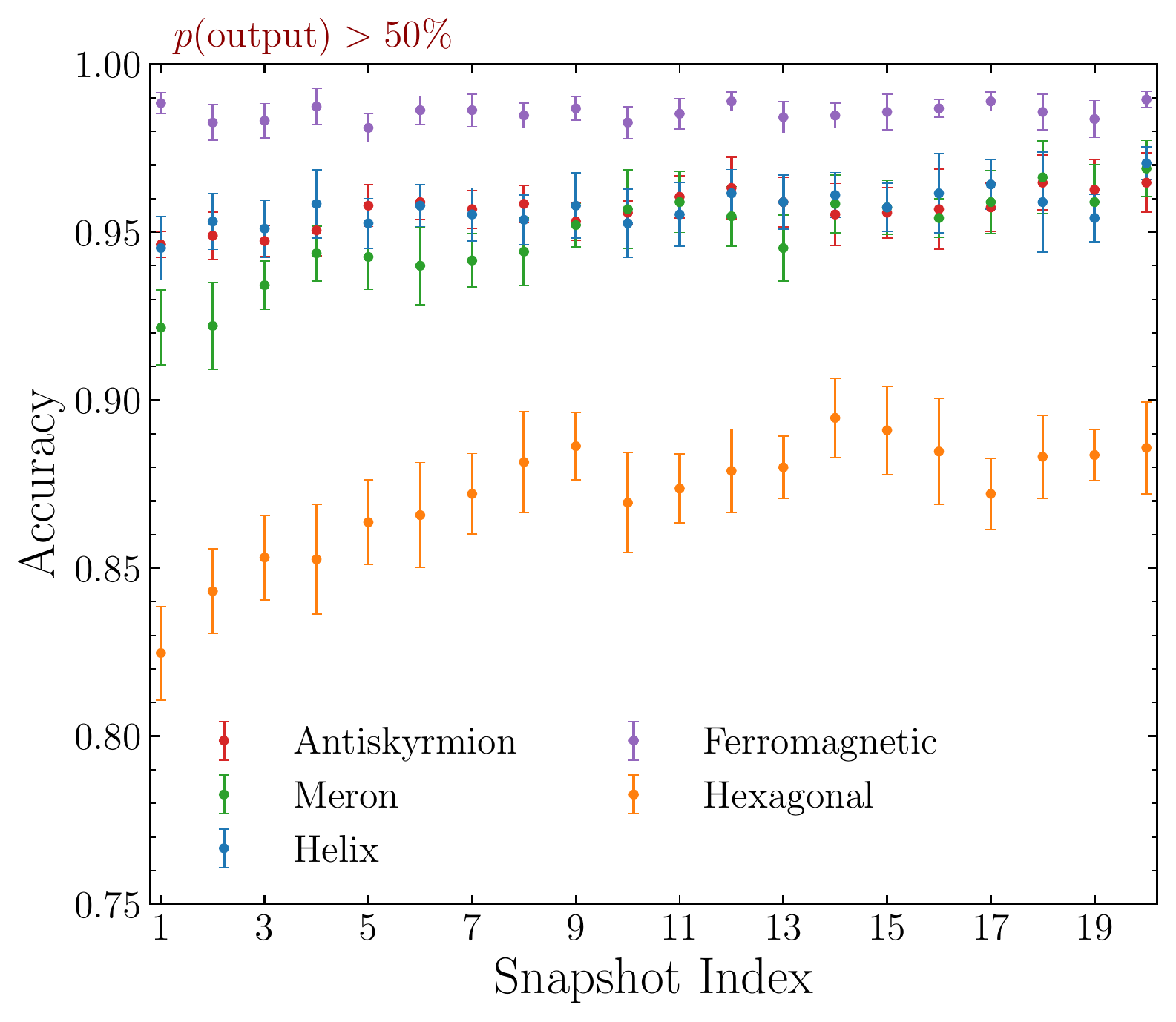}
  \caption{The accuracy of the CNN predictions as a function of the index of the snapshot used. Each point represents the mean accuracy of ten independent training with one standard deviation in the outcome. Snapshots are tagged if and only if output probability is greater than 50\%.}
  \label{fig:accuracies}
\end{figure}

In Fig.~\ref{fig:accuracies}, we show how the final accuracy changes as the network is trained on different snapshots from the simulations. We have performed the training independently for each snapshot, with the same architecture for all of them. In order to estimate the bias originating from the optimisation algorithm, the training is repeated 10 times for every snapshot with independent initiations. We present the results as central points representing the mean accuracy from the 10 runs, together with the error bars showing one standard deviation from the mean. Each label is assigned if the corresponding network output is above a threshold of 50\%\footnote{The network output does not refer to a specific certainty measure, where 90\% output does not ensure that the network is absolutely certain with the labelling. Here we used the network output as a threshold for a complex observable designed by the network.}. The colours distinguish the accuracies for the different labels defined in Table~\ref{tab:labels} where \texttt{antiskyrmion}, \texttt{bimeron}, \texttt{helix}, \texttt{ferromagnetic} and \texttt{hexagonal} are represented by red, green, blue, purple and orange colours.

As Fig.~\ref{fig:accuracies} shows, the performance of the network for most labels improves as it is trained on snapshots generated with longer Monte Carlo times until the snapshot index is around 5--10. This automatically identifies the point at which the simulation can be stopped: at snapshot 10, the network can already predict each label with the same accuracy as for the final one.

For fixed snapshot index, there is a hierarchy in the network's performance for different classes. With its arrangement of aligned spins, the ferromagnetic label is the easiest to identify, giving an accuracy of around 99\% for all snapshots. The antiskyrmion, bimeron and helix features are next, all of them having a similar final accuracy of approximately 95--96\%. The detection of the hexagonal arrangement is not so precise, with a maximal accuracy close to 90\%. This is to be expected since it is the only label that does not depend on the local features of the images but their global structure.

\begin{figure}
  \centering
  \vspace{10pt}
  \includegraphics[width=0.8\textwidth]{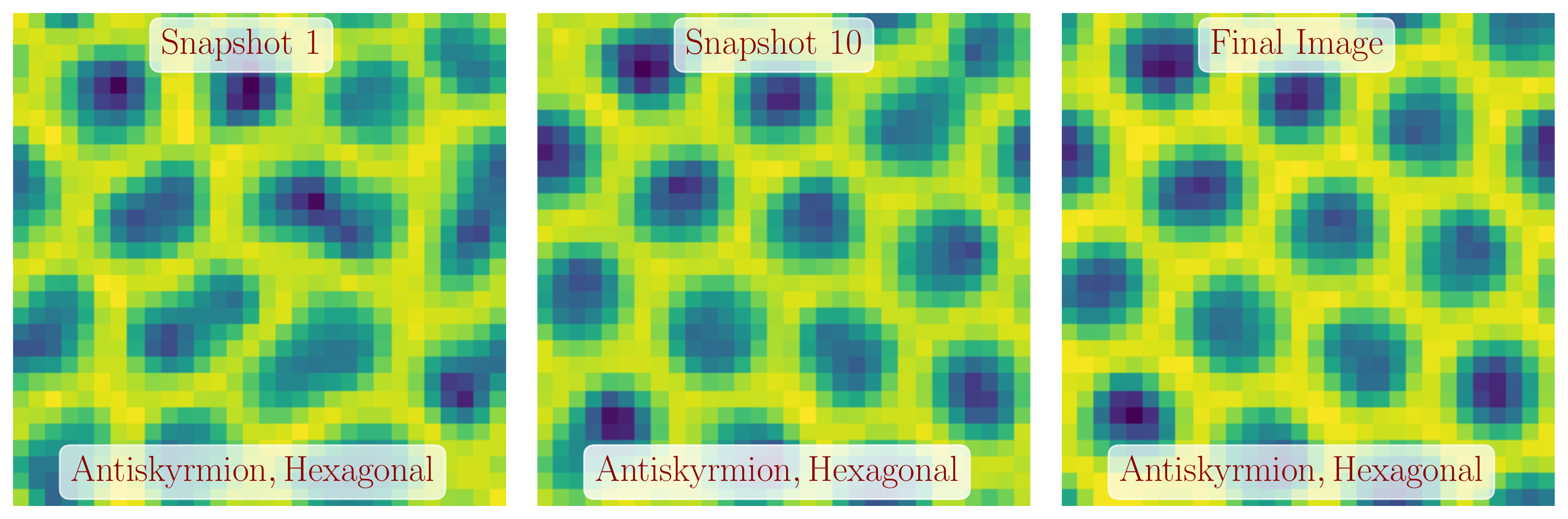}
  \includegraphics[width=0.8\textwidth]{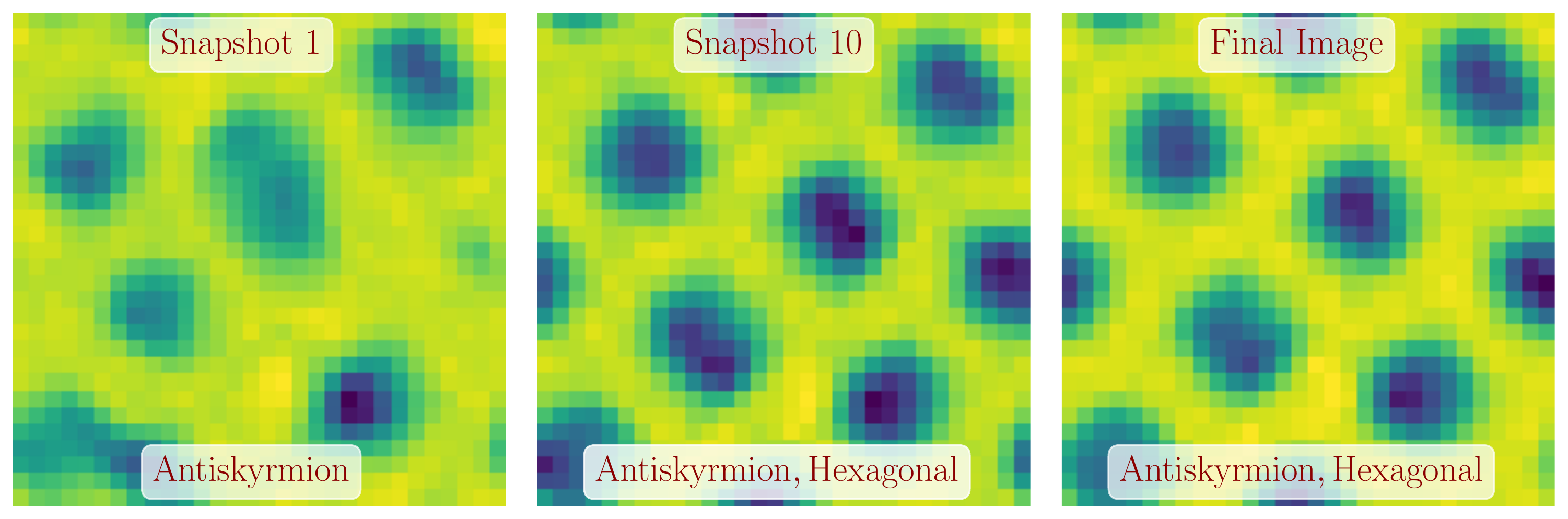}
  \includegraphics[width=0.8\textwidth]{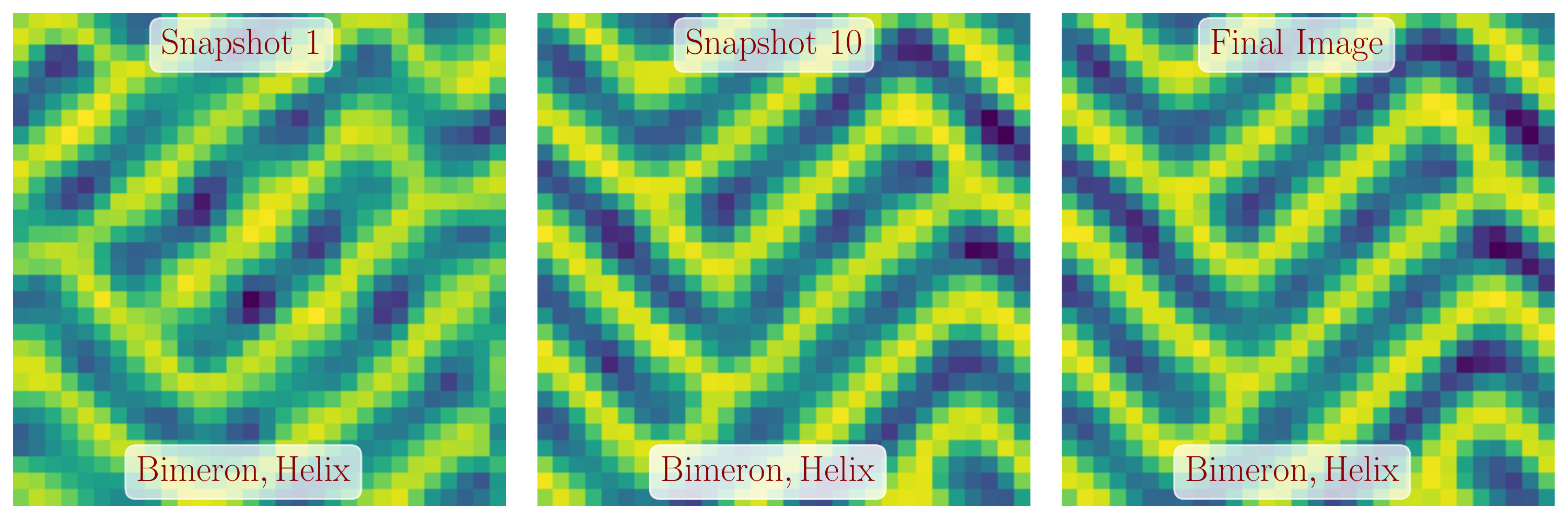}
  \caption{Examples of snapshots from different simulations. Each row have been divided in three columns for first and tenth snapshot and final image. The bottom label in the final image shows the truth label of the image and the rest shows the mean output of the CNN with output probability greater than 50\%.}
  \label{fig:examples}
\end{figure}

We illustrate the evolution of the network predictions for three samples in Fig.~\ref{fig:examples}. The first row shows the remarkable performance of the network for the early snapshots: to the naked eye, the first snapshot looks like it contains antiskyrmions with strong deformations, which could be even tagged as bimerons, and no hexagonal structure. Intuitively, it is not clear from this what the evolution will be. Despite this, the network can predict that the final phase will only contain antiskyrmions, not bimerons, and be arranged in a hexagonal lattice. Finally, we stress the importance of training the network on the snapshot index it is expected to be used: the network for the final image, which achieves high accuracy in its domain, performs worse than the one specialised in the first snapshot index here, giving the incorrect \texttt{[antiskyrmion, bimeron]} prediction.

In the second row of Fig.~\ref{fig:examples}, one can see the first snapshot with structures that the noise makes hard to classify by hand. One of them appears to be a wall of spins on the bottom left of the image, which could be misidentified as a bimeron being formed. This ends up disappearing in later snapshots. The network correctly predicts this by attaching a single \texttt{[antiskyrmion]} label to it. However, in this case, one needs to wait until snapshot 10 for the network to predict the hexagonal arrangement of antiskyrmions. Finally, in the third row of Fig.~\ref{fig:examples}, we show an example with bimeron/helical features which is correctly classified from the beginning, despite the noise that is present there.

As an example of the practical application of our approach, we use it to generate a phase diagram from the results of simulations that are stopped at early times and compare its application to the results of the full simulations. First, we generate samples in the same way as the training set, but this time for an equally-spaced grid of points in $(T, B)$ space, with $K = \tan(2\pi / 7) \simeq 1.25$. The temperature $T$ varies between 0.05 to 1.25, in steps of 0.05, while the external magnetic field $B$ goes from 0 to 0.4, in steps of 0.02. Then, we apply the previously-trained networks for snapshot indexes 6 and 20 to the corresponding snapshots and generate a phase diagram. To have smoother borders for each region in both diagrams, we take the average of the predictions for each $2 \times 2$ block in the $(T, B)$ grid. We then assign a colour/label to each point as follows:
\begin{itemize}
 \item Points with a network output above 0.5 for the antiskyrmion label are coloured in red and labelled ``Antiskyrmion''. We use a ligher red color when the output for the ferromagnetic label is above 0.5 and an even lighter one when it becomes larger than 0.75. The mixture of antiskyrmions and spin-aligned (ferromagnetic) regions points to the existence of an ``Antiskyrmion gas'' here, as discussed below.
 \item Points whose only network output above 0.5 is the ferromagnetic one are labelled ``Ferromagnetic'' and assigned a grey colour.
 \item Similarly, those points whose only network output above 0.5 is the helical one are labelled ``Helical'' and assigned a blue colour.
 \item Finally, we select a lower threshold of 0.05 for the bimeron output. If this is reached, independent of other labels, the corresponding point is labelled ``Bimeron'' and uses a light salmon colour. An slighly ligher color is used if the output reaches the value 0.25. This is done to show a region where bimerons might appear, in conjunction with antiskyrmions, helical-phase walls or both. The choice of a lower threshold is made to show this region more clearly. For 0.5, it is smaller, and it almost disappears at snapshot index 19.
\end{itemize}
The points that do not fit any of these criteria are left in white colour, signalling that they could not be classified confidently in any of these phases.

\begin{figure}
  \centering
  \includegraphics[width=0.55\textwidth]{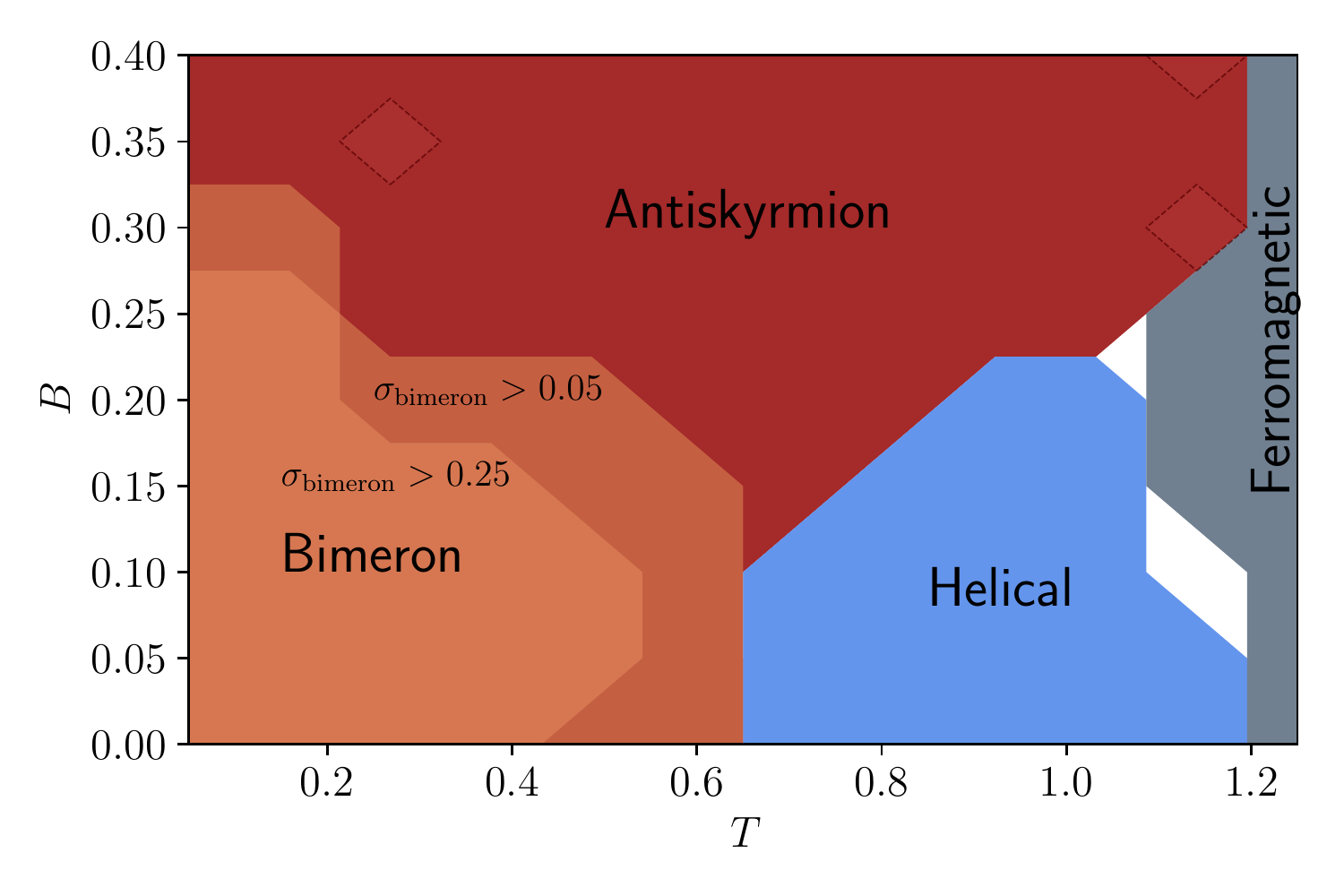}
  \includegraphics[width=0.55\textwidth]{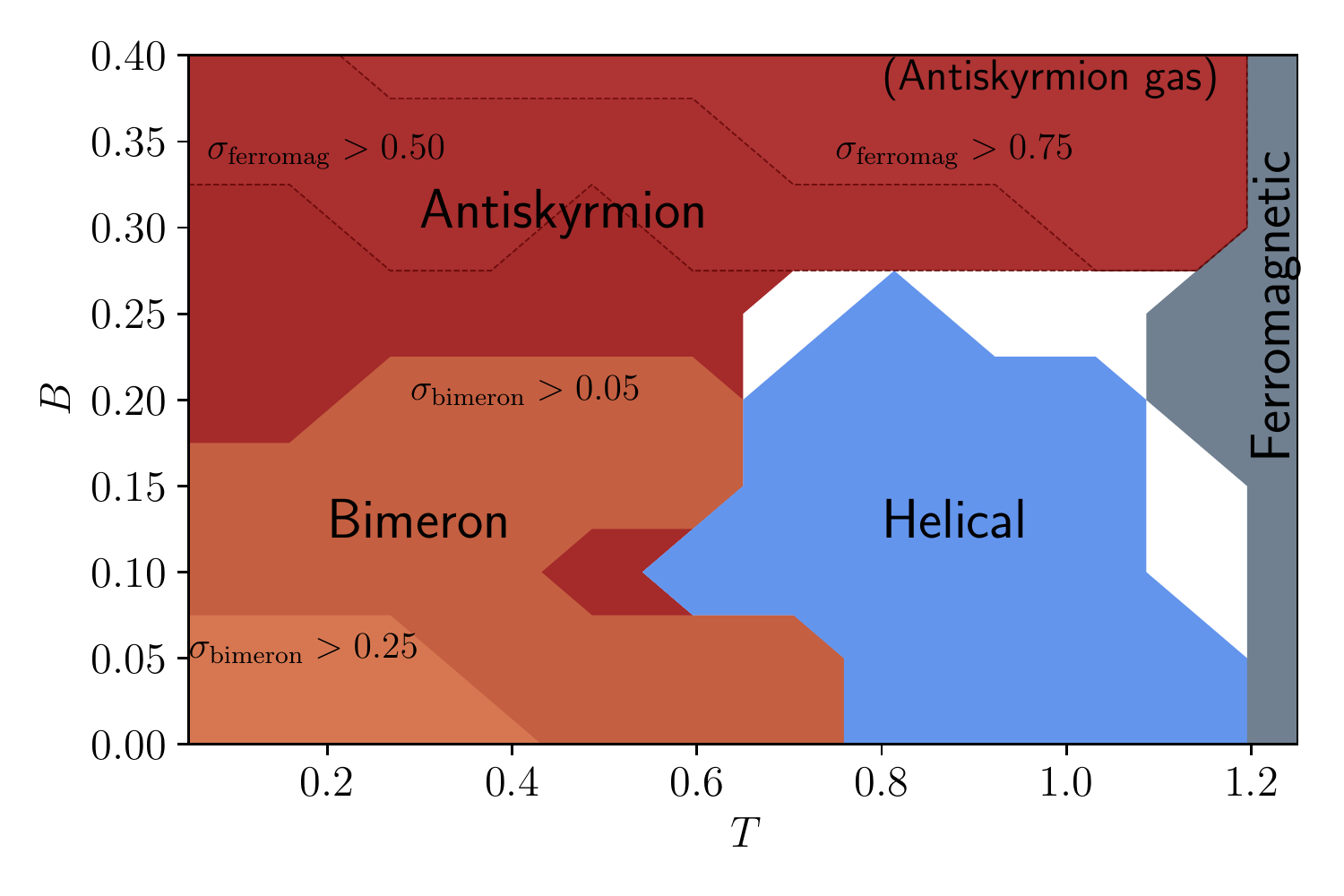}
  \includegraphics[width=0.55\textwidth]{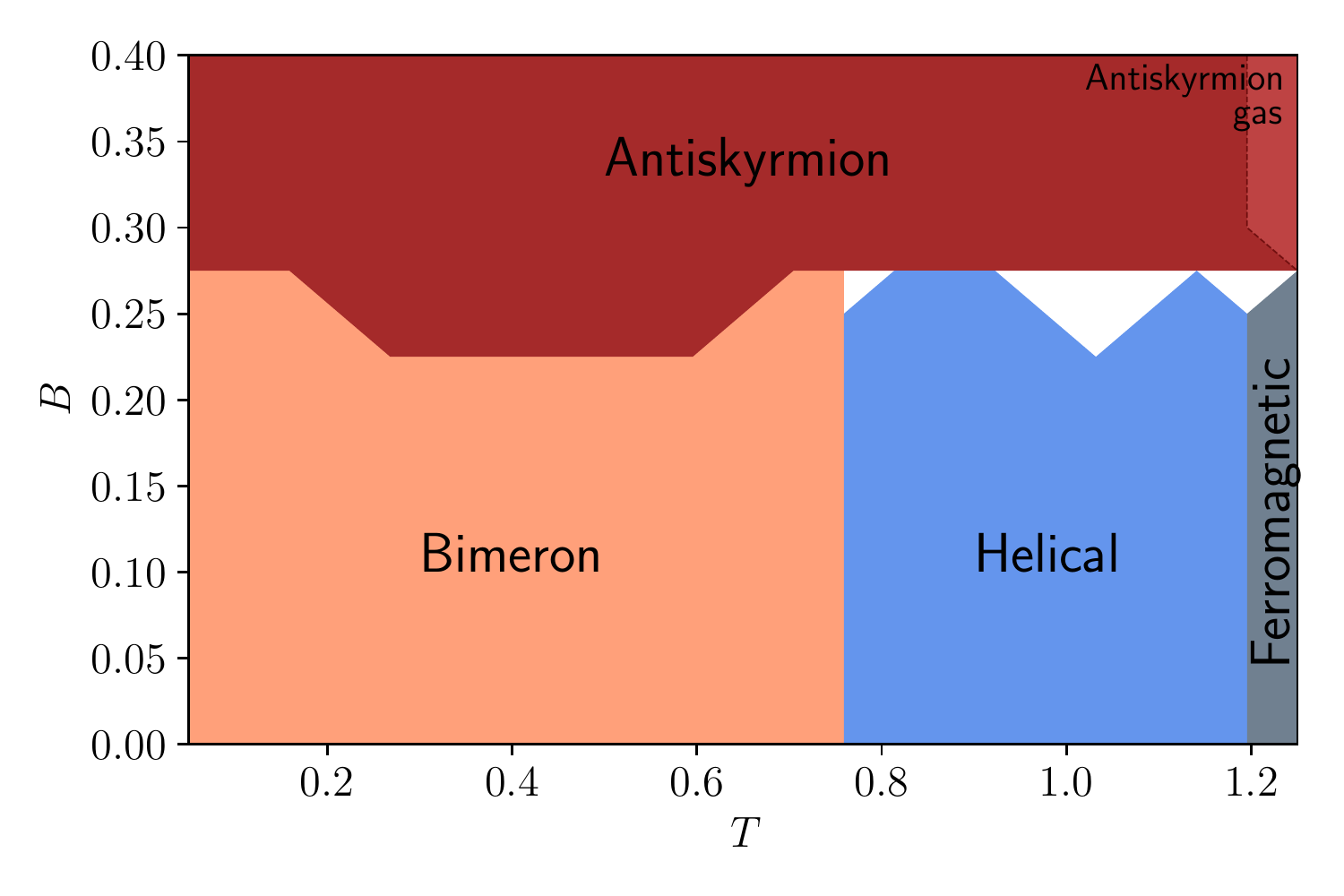}
  \caption{Left: phase diagram for a constant schedule, obtained from short (snapshot 6) simulations followed by network prediction. Left: the same diagram, using instead long (snapshot 20) simulations. The sub-regions labeled $\sigma_i > t$ contain all the points in which the output of the network for class $i$ is above the given $t$. Right: the same diagram, labeled by hand using the images from the long simulations. The main regions of a typical phase diagram (pure helical, pure ferromagnetic, and the complete region where antiskyrmions are found), are identified correctly from early stages of the simulation. The network for longer simulation times predicts a ferromagnetic label for the upper-right corner, pointing to the existence of an antiskyrmion gas. This is not found for shorter times.}
  \label{fig:phase-diagram}
\end{figure}

We show the resulting phase diagrams in Fig.~\ref{fig:phase-diagram}. We also include a phase diagram generated by labeling the snapshot-20 data by hand. Since the schedule consists of thermalising at a constant value of $T$ and $B$, starting from a random lattice configuration, these diagrams correspond to what may be obtained experimentally by rapidly cooling from a high temperature directly into the target point. Perhaps surprisingly, the bimeron and antiskyrmion structures survive at low temperatures in this context. This due to the specific schedule being used, in which these structures quickly form and become frozen because of the small perturbations generated by the small value of the temperature. With other schedules, such as zero-field cooling, high-field cooling, or even a slower direct cooling, this effect is likely to disappear. Despite this, the effect is not unphysical: it could be reproduced experimentally by quickly lowering the temperature of a hot material.

We first notice the following differences between the network-generated phase diagrams for the two selected snapshots: the shrinking of the ``Bimeron'' region and the appearance of the ``Antiskyrmion gas'' one. The shrinking of the ``Bimeron'' region can be explained by the presence in the early stages of structures that are classified as bimerons, which later disappear in favour of antiskyrmions. This is partially mitigated by training the network on the early snapshots, as shown in some of the examples in Fig.~\ref{fig:examples}, but the effect persists to a smaller degree. By comparing with the manually-generated diagram, we find that the snapshot-20 network slightly suffers from the opposite effect: it produces a small output value for the \texttt{bimeron} class even when bimerons are present. However, the diagram shows that this can be compensated for by using a lower threshold value for this class.

Another feature that can only be found at larger Monte Carlo times is the mixed antiskyrmion-ferromagnetic states, which is not detectable at early stages due to noise. The co-existence of regions with aligned spins (ferromagnetic) and antiskyrmions suggests an ``Antiskyrmion gas'' in which the antiskyrmions are not compactly packed. By directly inspecting the data, we confirm that such states exist in the small top-right region shown in the manually-labeld phase diagram. In this case, the simulations need to be run for the entire time we considered to resolve this detail.

However, the rest of both diagrams approximately agree. Concretely, the regions for the helical and ferromagnetic phase and the regions where antiskyrmions are detected are similar. This means that to identify where these features emerge, one only needs to simulate snapshot 6. Since these are the most critical features, and the only ones present in a typical phase diagram, running short simulations and using the CNN to predict a final state is a viable option for many applications.

Finally, as a simple test for the claim that the method presented here would work similarly for skyrmions as it does for antiskyrmions, because of the similarity in their $z$-component distribution, we generate 100 random samples with the $T$, $B$, $K$ parameters in the ranges considered above, but with DM interactions having the structure that generates Bloch skyrmions~\cite{Criado:2021gzp}. We then apply the CNN that we trained on the final-snapshot antiskyrmion samples to this dataset. Using the antiskyrmion label as a proxy for the identification of skyrmions, we find an accuracy of $86\%$. We take this as a strong indication that the performance of a CNN trained directly on the skyrmion samples would be comparable to the one we found for antiskyrmions.

\section{Conclusions}
\label{sec:conclusions}

We have shown that a multi-label machine learning approach allows us to identify complex structures and mixed states in the results of 3D Monte Carlo simulations of spin lattices with DM interactions. We have trained a CNN using 2D lattice slices to detect antiskyrmions, bimerons, helical-phase walls, regions with ferromagnetic arrangements of spins, and hexagonal lattices of antiskyrmions. We have only used the $z$ component of the lattice spins, which is similar among antiskyrmions and Bloch/Neel skyrmions. Although we have applied this approach to a version of the DM interactions that supports antiskyrmions, it would perform similarly for skyrmions of both types.

In addition to directly identifying these features, we have used this framework to predict their emergence from the early stages of the simulation. The CNNs trained on the first few snapshots, with labels given by the final configuration, can predict final features even in cases in which it is not intuitively apparent from an inspection by the naked eye of the corresponding images.

One of the applications of the early-snapshot CNNs is in shortening simulation times. Thanks to them, one can stop the simulation at early stages and predict the relevant features. As an example, we have constructed $(T, B)$ phase diagrams for snapshots 6 and 20 with $K = \tan(2\pi / 7)$. Although long simulation times are needed to resolve fine details involving mixtures of bimerons and antiskyrmions or whether antiskyrmions are in a crystal or a gas phase, we find that the main phases, antiskyrmion, helical, and ferromagnetic, can be detected with significantly shorter simulations.

\bibliographystyle{elsarticle-num}
\bibliography{references}

\begin{thebibliography}{10}
\expandafter\ifx\csname url\endcsname\relax
  \def\url#1{\texttt{#1}}\fi
\expandafter\ifx\csname urlprefix\endcsname\relax\def\urlprefix{URL }\fi
\expandafter\ifx\csname href\endcsname\relax
  \def\href#1#2{#2} \def\path#1{#1}\fi

\bibitem{Dzyaloshinskii:1958}
I.~Dzyaloshinsky,
  \href{http://www.sciencedirect.com/science/article/pii/0022369758900763}{A
  thermodynamic theory of weak ferromagnetism of antiferromagnetics}, Journal
  of Physics and Chemistry of Solids 4~(4) (1958) 241 -- 255.
\newblock \href {https://doi.org/https://doi.org/10.1016/0022-3697(58)90076-3}
  {\path{doi:https://doi.org/10.1016/0022-3697(58)90076-3}}.
\newline\urlprefix\url{http://www.sciencedirect.com/science/article/pii/0022369758900763}

\bibitem{Moriya:1960zz}
T.~Moriya, {Anisotropic Superexchange Interaction and Weak Ferromagnetism},
  Phys. Rev. 120 (1960) 91--98.
\newblock \href {https://doi.org/10.1103/PhysRev.120.91}
  {\path{doi:10.1103/PhysRev.120.91}}.

\bibitem{Skyrme:1962vh}
T.~H.~R. Skyrme, {A Unified Field Theory of Mesons and Baryons}, Nucl. Phys. 31
  (1962) 556--569.
\newblock \href {https://doi.org/10.1016/0029-5582(62)90775-7}
  {\path{doi:10.1016/0029-5582(62)90775-7}}.

\bibitem{Bogdanov:1989}
A.~N. Bogdanov, D.~Yablonskii, Thermodynamically stable ``vortices'' in
  magnetically ordered crystals. the mixed state of magnets, Zh. Eksp. Teor.
  Fiz 95~(1) (1989) 178.

\bibitem{Muhlbauer:2009}
S.~Muhlbauer, B.~Binz, F.~Jonietz, C.~Pfleiderer, A.~Rosch, A.~Neubauer,
  R.~Georgii, P.~Boni,
  \href{http://dx.doi.org/10.1126/science.1166767}{Skyrmion lattice in a chiral
  magnet}, Science 323~(5916) (2009) 915?919.
\newblock \href {https://doi.org/10.1126/science.1166767}
  {\path{doi:10.1126/science.1166767}}.
\newline\urlprefix\url{http://dx.doi.org/10.1126/science.1166767}

\bibitem{Munzer:2009var}
W.~M\"unzer, et~al., {Skyrmion lattice in the doped semiconductor
  Fe$_{1-x}$Co$_x$Si}, Phys. Rev. B 81~(4) (2010) 041203.
\newblock \href {http://arxiv.org/abs/0903.2587} {\path{arXiv:0903.2587}},
  \href {https://doi.org/10.1103/PhysRevB.81.041203}
  {\path{doi:10.1103/PhysRevB.81.041203}}.

\bibitem{yu2010real}
X.~Yu, Y.~Onose, N.~Kanazawa, J.~H. Park, J.~Han, Y.~Matsui, N.~Nagaosa,
  Y.~Tokura, Real-space observation of a two-dimensional skyrmion crystal,
  Nature 465~(7300) (2010) 901--904.
\newblock \href {https://doi.org/10.1038/nature09124}
  {\path{doi:10.1038/nature09124}}.

\bibitem{yu2011near}
X.~Yu, N.~Kanazawa, Y.~Onose, K.~Kimoto, W.~Zhang, S.~Ishiwata, Y.~Matsui,
  Y.~Tokura, Near room-temperature formation of a skyrmion crystal in
  thin-films of the helimagnet fege, Nature materials 10~(2) (2011) 106--109.
\newblock \href {https://doi.org/10.1038/nmat2916}
  {\path{doi:10.1038/nmat2916}}.

\bibitem{tokunaga2015new}
Y.~Tokunaga, X.~Yu, J.~White, H.~M. R{\o}nnow, D.~Morikawa, Y.~Taguchi,
  Y.~Tokura, A new class of chiral materials hosting magnetic skyrmions beyond
  room temperature, Nature communications 6~(1) (2015) 1--7.
\newblock \href {https://doi.org/10.1038/ncomms8638}
  {\path{doi:10.1038/ncomms8638}}.

\bibitem{woo2016observation}
S.~{Woo}, K.~{Litzius}, B.~{Kr{\"u}ger}, M.-Y. {Im}, L.~{Caretta},
  K.~{Richter}, M.~{Mann}, A.~{Krone}, R.~M. {Reeve}, M.~{Weigand},
  P.~{Agrawal}, I.~{Lemesh}, M.-A. {Mawass}, P.~{Fischer}, M.~{Kl{\"a}ui},
  G.~S.~D. {Beach}, {Observation of room-temperature magnetic skyrmions and
  their current-driven dynamics in ultrathin metallic ferromagnets}, Nature
  Materials 15~(5) (2016) 501--506.
\newblock \href {http://arxiv.org/abs/1502.07376} {\path{arXiv:1502.07376}},
  \href {https://doi.org/10.1038/nmat4593} {\path{doi:10.1038/nmat4593}}.

\bibitem{fujima2017thermodynamically}
Y.~{Fujima}, N.~{Abe}, Y.~{Tokunaga}, T.~{Arima}, {Thermodynamically stable
  skyrmion lattice at low temperatures in a bulk crystal of lacunar spinel
  GaV$_{4}$Se$_{8}$}, \prb 95~(18) (2017) 180410.
\newblock \href {https://doi.org/10.1103/PhysRevB.95.180410}
  {\path{doi:10.1103/PhysRevB.95.180410}}.

\bibitem{koshibae2016theory}
W.~Koshibae, N.~Nagaosa, Theory of antiskyrmions in magnets, Nature
  communications 7~(1) (2016) 1--8.

\bibitem{hoffmann2017antiskyrmions}
M.~{Hoffmann}, B.~{Zimmermann}, G.~P. {M{\"u}ller}, D.~{Sch{\"u}rhoff}, N.~S.
  {Kiselev}, C.~{Melcher}, S.~{Bl{\"u}gel}, {Antiskyrmions stabilized at
  interfaces by anisotropic Dzyaloshinskii-Moriya interactions}, Nature
  Communications 8 (2017) 308.
\newblock \href {http://arxiv.org/abs/1702.07573} {\path{arXiv:1702.07573}},
  \href {https://doi.org/10.1038/s41467-017-00313-0}
  {\path{doi:10.1038/s41467-017-00313-0}}.

\bibitem{huang2017stabilization}
S.~{Huang}, C.~{Zhou}, G.~{Chen}, H.~{Shen}, A.~K. {Schmid}, K.~{Liu}, Y.~{Wu},
  {Stabilization and current-induced motion of antiskyrmion in the presence of
  anisotropic Dzyaloshinskii-Moriya interaction}, \prb 96~(14) (2017) 144412.
\newblock \href {http://arxiv.org/abs/1709.07156} {\path{arXiv:1709.07156}},
  \href {https://doi.org/10.1103/PhysRevB.96.144412}
  {\path{doi:10.1103/PhysRevB.96.144412}}.

\bibitem{camosi2018micromagnetics}
L.~{Camosi}, N.~{Rougemaille}, O.~{Fruchart}, J.~{Vogel}, S.~{Rohart},
  {Micromagnetics of antiskyrmions in ultrathin films}, \prb 97~(13) (2018)
  134404.
\newblock \href {http://arxiv.org/abs/1712.04743} {\path{arXiv:1712.04743}},
  \href {https://doi.org/10.1103/PhysRevB.97.134404}
  {\path{doi:10.1103/PhysRevB.97.134404}}.

\bibitem{kovalev2018skyrmions}
A.~A. {Kovalev}, S.~{Sandhoefner}, {Skyrmions and antiskyrmions in
  quasi-two-dimensional magnets}, Frontiers in Physics 6 (2018) 98.
\newblock \href {https://doi.org/10.3389/fphy.2018.00098}
  {\path{doi:10.3389/fphy.2018.00098}}.

\bibitem{bottcher2018b}
M.~{B{\"o}ttcher}, S.~{Heinze}, S.~{Egorov}, J.~{Sinova}, B.~{Dup{\'e}}, {B-T
  phase diagram of Pd/Fe/Ir(111) computed with parallel tempering Monte Carlo},
  New Journal of Physics 20~(10) (2018) 103014.
\newblock \href {http://arxiv.org/abs/1707.01708} {\path{arXiv:1707.01708}},
  \href {https://doi.org/10.1088/1367-2630/aae282}
  {\path{doi:10.1088/1367-2630/aae282}}.

\bibitem{jena2020elliptical}
J.~{Jena}, B.~{G{\"o}bel}, T.~{Ma}, V.~{Kumar}, R.~{Saha}, I.~{Mertig},
  C.~{Felser}, S.~S.~P. {Parkin}, {Elliptical Bloch skyrmion chiral twins in an
  antiskyrmion system}, Nature Communications 11 (2020) 1115.
\newblock \href {https://doi.org/10.1038/s41467-020-14925-6}
  {\path{doi:10.1038/s41467-020-14925-6}}.

\bibitem{yu2021magnetic}
X.~Yu, Magnetic imaging of various topological spin textures and their
  dynamics, Journal of Magnetism and Magnetic Materials 539 (2021) 168332.

\bibitem{Fert:2013}
A.~{Fert}, V.~{Cros}, J.~{Sampaio}, {Skyrmions on the track}, Nature
  Nanotechnology 8~(3) (2013) 152--156.
\newblock \href {https://doi.org/10.1038/nnano.2013.29}
  {\path{doi:10.1038/nnano.2013.29}}.

\bibitem{tomasello2014strategy}
R.~{Tomasello}, E.~{Martinez}, R.~{Zivieri}, L.~{Torres}, M.~{Carpentieri},
  G.~{Finocchio}, {A strategy for the design of skyrmion racetrack memories},
  Scientific Reports 4 (2014) 6784.
\newblock \href {http://arxiv.org/abs/1409.6491} {\path{arXiv:1409.6491}},
  \href {https://doi.org/10.1038/srep06784} {\path{doi:10.1038/srep06784}}.

\bibitem{song2020skyrmion}
K.~M. Song, J.-S. Jeong, B.~Pan, X.~Zhang, J.~Xia, S.~Cha, T.-E. Park, K.~Kim,
  S.~Finizio, J.~Raabe, et~al., Skyrmion-based artificial synapses for
  neuromorphic computing, Nature Electronics 3~(3) (2020) 148--155.
\newblock \href {http://arxiv.org/abs/1907.00957} {\path{arXiv:1907.00957}},
  \href {https://doi.org/10.1038/s41928-020-0385-0}
  {\path{doi:10.1038/s41928-020-0385-0}}.

\bibitem{pinna2020reservoir}
D.~{Pinna}, G.~{Bourianoff}, K.~{Everschor-Sitte}, {Reservoir Computing with
  Random Skyrmion Textures}, Physical Review Applied 14~(5) (2020) 054020.
\newblock \href {https://doi.org/10.1103/PhysRevApplied.14.054020}
  {\path{doi:10.1103/PhysRevApplied.14.054020}}.

\bibitem{zazvorka2019thermal}
J.~{Z{\'a}zvorka}, F.~{Jakobs}, D.~{Heinze}, N.~{Keil}, S.~{Kromin},
  S.~{Jaiswal}, K.~{Litzius}, G.~{Jakob}, P.~{Virnau}, D.~{Pinna},
  K.~{Everschor-Sitte}, L.~{R{\'o}zsa}, A.~{Donges}, U.~{Nowak},
  M.~{Kl{\"a}ui}, {Thermal skyrmion diffusion used in a reshuffler device},
  Nature Nanotechnology 14~(7) (2019) 658--661.
\newblock \href {https://doi.org/10.1038/s41565-019-0436-8}
  {\path{doi:10.1038/s41565-019-0436-8}}.

\bibitem{buhrandt2013skyrmion}
S.~Buhrandt, L.~Fritz, Skyrmion lattice phase in three-dimensional chiral
  magnets from monte carlo simulations, Physical Review B 88~(19) (2013)
  195137.

\bibitem{Criado:2021gzp}
J.~C. Criado, P.~D. Hatton, S.~Schenk, M.~Spannowsky, L.~A. Turnbull,
  {Simulating magnetic antiskyrmions on the lattice} (9 2021).
\newblock \href {http://arxiv.org/abs/2109.15020} {\path{arXiv:2109.15020}}.

\bibitem{PhysRevB.98.174411}
I.~A. Iakovlev, O.~M. Sotnikov, V.~V. Mazurenko,
  \href{https://link.aps.org/doi/10.1103/PhysRevB.98.174411}{Supervised
  learning approach for recognizing magnetic skyrmion phases}, Phys. Rev. B 98
  (2018) 174411.
\newblock \href {https://doi.org/10.1103/PhysRevB.98.174411}
  {\path{doi:10.1103/PhysRevB.98.174411}}.
\newline\urlprefix\url{https://link.aps.org/doi/10.1103/PhysRevB.98.174411}

\bibitem{salcedo2020deep}
J.~Salcedo-Gallo, C.~Galindo-Gonz{\'a}lez, E.~Restrepo-Parra, Deep learning
  approach for image classification of magnetic phases in chiral magnets,
  Journal of Magnetism and Magnetic Materials 501 (2020) 166482.

\bibitem{singh2019application}
V.~K. Singh, J.~H. Han, Application of machine learning to two-dimensional
  dzyaloshinskii-moriya ferromagnets, Physical Review B 99~(17) (2019) 174426.

\bibitem{PhysRevB.105.214423}
F.~A. G\'omez~Albarrac\'{\i}n, H.~D. Rosales,
  \href{https://link.aps.org/doi/10.1103/PhysRevB.105.214423}{Machine learning
  techniques to construct detailed phase diagrams for skyrmion systems}, Phys.
  Rev. B 105 (2022) 214423.
\newblock \href {https://doi.org/10.1103/PhysRevB.105.214423}
  {\path{doi:10.1103/PhysRevB.105.214423}}.
\newline\urlprefix\url{https://link.aps.org/doi/10.1103/PhysRevB.105.214423}

\bibitem{perzhu2020computer}
A.~Perzhu, E.~Vasiliev, D.~Kapitan, A.~Rybin, A.~Korol, K.~Nefedev, V.~Kapitan,
  Computer simulation of skyrmions on a square lattice (2020).

\bibitem{kapitan2021numerical}
V.~Kapitan, E.~Vasiliev, A.~Perzhu, D.~Kapitan, A.~Rybin, A.~Korol,
  K.~Soldatov, Y.~Shevchenko, Numerical simulation of magnetic skyrmions on
  flat lattices, AIP Advances 11~(1) (2021) 015041.

\bibitem{matthies2022topological}
T.~Matthies, A.~F. Sch{\"a}ffer, T.~Posske, R.~Wiesendanger, E.~Y. Vedmedenko,
  Topological characterization of dynamic chiral magnetic textures using
  machine learning, arXiv preprint arXiv:2201.01629 (2022).

\bibitem{kawaguchi2021determination}
M.~Kawaguchi, K.~Tanabe, K.~Yamada, T.~Sawa, S.~Hasegawa, M.~Hayashi,
  Y.~Nakatani, Determination of the dzyaloshinskii-moriya interaction using
  pattern recognition and machine learning, npj Computational Materials 7~(1)
  (2021) 1--7.

\bibitem{wang2021learning}
W.~Wang, Z.~Wang, Y.~Zhang, B.~Sun, K.~Xia, Learning order parameters from
  videos of skyrmion dynamical phases with neural networks, Physical Review
  Applied 16~(1) (2021) 014005.

\bibitem{GOBEL20211}
B.~Göbel, I.~Mertig, O.~A. Tretiakov,
  \href{https://www.sciencedirect.com/science/article/pii/S0370157320303525}{Beyond
  skyrmions: Review and perspectives of alternative magnetic quasiparticles},
  Physics Reports 895 (2021) 1--28, beyond skyrmions: Review and perspectives
  of alternative magnetic quasiparticles.
\newblock \href {https://doi.org/https://doi.org/10.1016/j.physrep.2020.10.001}
  {\path{doi:https://doi.org/10.1016/j.physrep.2020.10.001}}.
\newline\urlprefix\url{https://www.sciencedirect.com/science/article/pii/S0370157320303525}

\bibitem{PhysRevB.83.100408}
M.~Ezawa, \href{https://link.aps.org/doi/10.1103/PhysRevB.83.100408}{Compact
  merons and skyrmions in thin chiral magnetic films}, Phys. Rev. B 83 (2011)
  100408.
\newblock \href {https://doi.org/10.1103/PhysRevB.83.100408}
  {\path{doi:10.1103/PhysRevB.83.100408}}.
\newline\urlprefix\url{https://link.aps.org/doi/10.1103/PhysRevB.83.100408}

\bibitem{rosales2022skyrmion}
H.~Rosales, F.~A.~G. Albarrac{\'\i}n, P.~Pujol, L.~D. Jaubert, A skyrmion fluid
  and bimeron glass emerging from competition with a chiral spin liquid, arXiv
  preprint arXiv:2202.06993 (2022).

\bibitem{Kingma2014AdamAM}
D.~P. Kingma, J.~Ba, Adam: A method for stochastic optimization, CoRR
  abs/1412.6980 (2014).

\end{thebibliography}

\end{document}